# Exploration of Gate Trench Module for Vertical GaN devices

M. Ruzzarin[1], K. Geens[2], M. Borga[2], H. Liang[2], S. You[2], B. Bakeroot[3], S. Decoutere[2], C. De Santi[1], A. Neviani[1], M. Meneghini[1], G. Meneghesso[1], E. Zanoni[1]

[1] University of Padova, Department of Information Engineering, Via Gradenigo 6/b, 35131, Padova, Italy
[2] imec, Leuven, Kapeldreef 75, 3001 Leuven, Belgium
[3] CMST, imec and Ghent University, Technologiepark 126, Ghent, 9052, Belgium

*Abstract –* The aim of this work is to present the optimization of the gate trench module for use in vertical GaN devices: we considered the impact of cleaning process of the etched surface of the gate trench, thickness of gate dielectric, and magnesium concentration of the p-GaN layer. The analysis was carried out by comparing the main DC parameters of devices that differ in surface cleaning process of the gate trench, gate dielectric thickness, and body layer doping. On the basis of experimental results, we report that: (i) a good cleaning process of the etched GaN surface of the gate trench is a key factor to enhance the device performance, (ii) a gate dielectric >35-nm $SiO_2$ results in a narrow distribution for DC characteristics, (iii) lowering the p-doping in the body layer improves the ON-resistance ($R_{ON}$). Gate capacitance measurements are performed to further confirm the results. Hypotheses on dielectric trapping/de-trapping mechanisms under positive and negative gate bias are reported.

## 1. Introduction

Semiconductor electronic devices for power applications require normally-OFF operation, low ON-resistance, low leakage current and high breakdown voltage in order to reduce power losses and maintain reduced costs with electronics scaling [1][2]. Wide-bandgap transistors based on gallium nitride (GaN) are promising candidates for these requirements, due to the physical properties of the semiconductor in terms of high-conductivity, high frequency, high power operation. The most common GaN power devices with a high breakdown voltage have been AlGaN/GaN-based lateral HEMTs. In order to achieve normally-OFF operations the 2DEG below the gate region must be depleted, for example by using a p-type GaN layer between gate metal and AlGaN barrier [3][4][5]. Currently this approach is the most widely used for lateral GaN devices and excellent performance has been demonstrated with kV-range breakdown voltages [6][7]. However, in order to achieve higher breakdown voltage (> 1.5 kV) substantial gate-drain distance is necessary reducing the effective current density and increasing the chip size and cost. Moreover, the lateral devices are limited by the high substrate leakage and surface trapping effects [8]. In order to address high power levels (10-100kW) required for example in the automotive field, for data centers and photovoltaic systems, researchers are exploring vertical GaN devices. Compared to lateral structures, vertical GaN devices have several advantages: (i) the capability to achieve high breakdown voltage and current levels without enlarging the chip size, (ii) the superior reliability gained by moving the peak electric field away from the surface into bulk devices [9], (iii) the easier thermal management [10]. Several device architectures have been proposed: Current Aperture Vertical Electron Transistor (CAVET) [11], Vertical Field Effect Transistor (VFET)[12] and GaN nanowire arrays [13][14]. Also, Vertical GaN trench MOSFETs [15][16] are promising candidates for next generation of power devices: the regrowth of AlGaN/GaN structures is not needed and normally-OFF operation is intrinsically achieved [17]. However, there are still open issues, which include the material growth and device processing, the improvement of channel mobility and the research towards low-cost fabrication [18].

For GaN trench MOSFETs, the formation of the gate trench is fundamental to achieve a good performance, however the impact of the process used in the trench etching tends to result in rough surfaces in the trench [19]. A good dry etch process, is therefore necessary to bring the roughness to a minimum. Removing impurities and residuals created during the dry etch process steps with a good cleaning process is then necessary

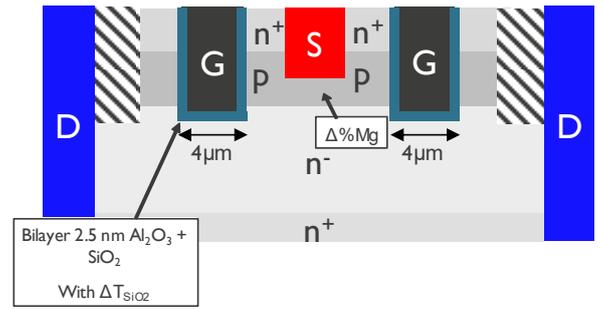

Fig. 1 Pictorial representation of the cross section of the GaN semi-vertical trench MOSFET. The devices tested in this work vary in cleaning process of the etched surface of the gate trench, $SiO_2$ thickness and p-doping level.

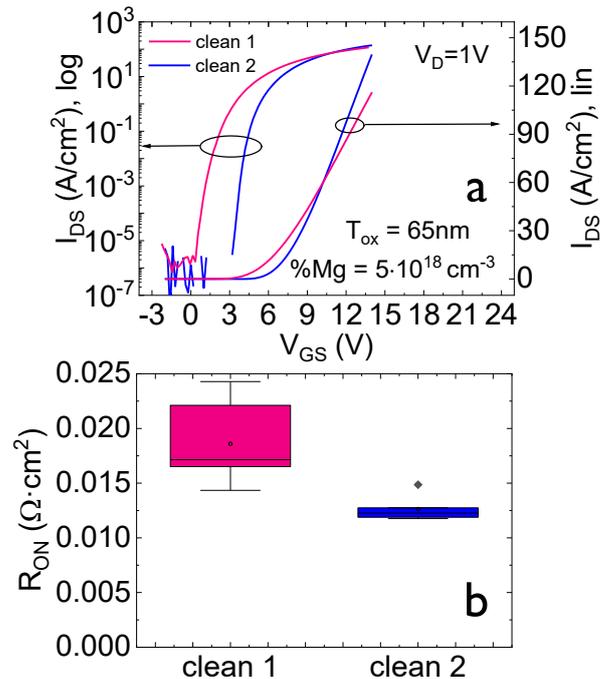

Fig. 2 $I_D V_{GS}$ comparison in semi-log and linear scale for 2 devices with different cleaning processes: clean 1 and clean 2 (a). Comparison of $R_{ON}$ measured for clean 1 and clean 2 (b).

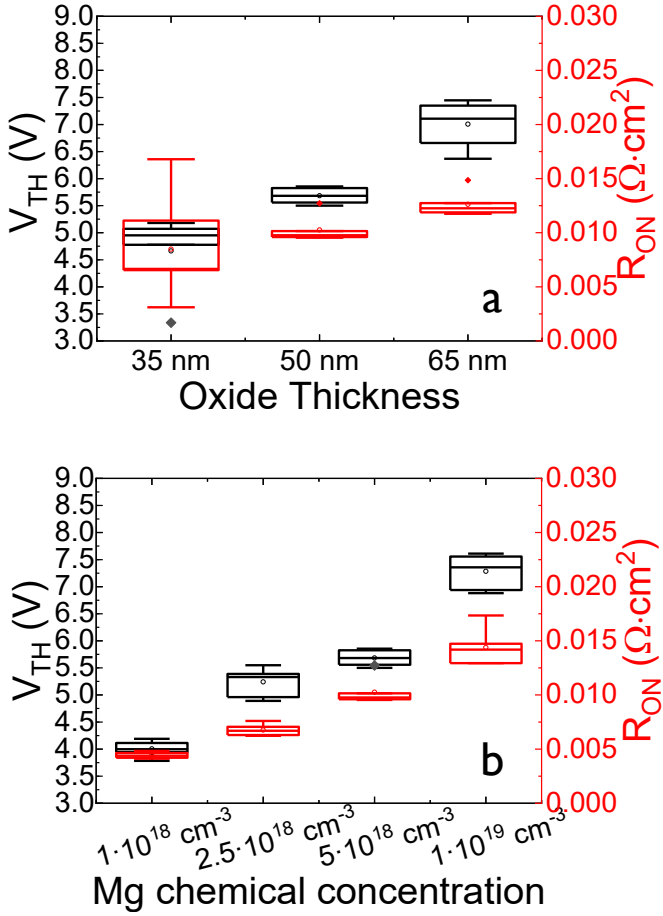

Fig. 3 Comparison of the threshold voltage and on-resistance.

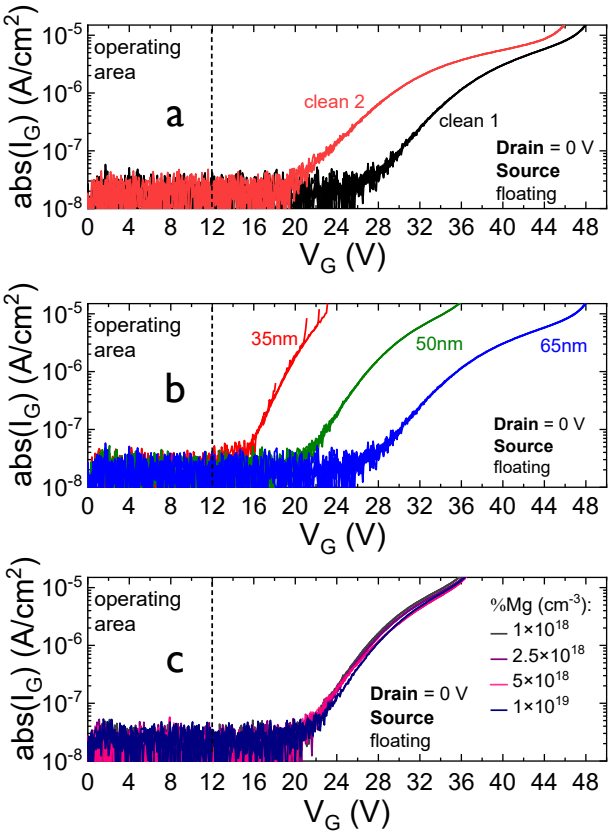

Fig. 4 Gate leakage current obtained for different cleaning processes (a), oxide thickness (b) and Mg chemical concentration (c).

[20][21][22]. Previous works demonstrated that a way to enhance the reliability of a device in the on state is to reduce the oxide field by increasing the gate oxide thickness. However, increasing the gate oxide thickness can increase the $R_{ON}$ [23][24]. From the ON-resistance model reported by [25], $R_{ON}$ is given by the sum of the channel resistance ($R_{MOS}$), and the series resistance.

$R_{MOS}$ is given by:

$$R_{MOS} = \frac{\frac{L}{W}}{\mu C_{ox}(V_G - I_{DS}R_{GS} - V_{FB})}$$

Under the assumption of varying only the dielectric thickness (the channel mobility $\mu$, the gate-source resistance $R_{GS}$, $I_{DS}$ and ($V_G$-$V_{FB}$) are supposed to be constants), since $C_{OX}$ is inversely proportional to $T_{OX}$, $R_{ON}$ depends on $T_{OX}$.

Another way to decrease the $R_{ON}$ is lowering the p-doping in the p-layer [26][27]. This work reports an extensive analysis of the impact of some of these factors (cleaning process, dielectric thickness, body doping) on the DC characteristics of GaN semi-vertical devices with an $Al_2O_3$ – $SiO_2$ stack as gate dielectric. Moreover, the gate instability under positive and negative biases is investigated and hypotheses on the trapping/detrapping mechanisms are given.

## 2. Experimental details

In this study, normally-OFF GaN-on-Si semi-vertical trench MOSFETs with different splits in the gate module are tested to analyze the impact of the gate trench on the device performance. The semi-vertical architecture is used as a test vehicle to develop optimal process modules for later implementation in a vertical device architecture. The gate dielectric consists of a bilayer [28] of 2.5 nm $Al_2O_3$ and $SiO_2$ with different thicknesses. The devices are grown on a silicon substrate. A schematic representation of the cross section of the device under test (DUT) is reported in Fig.1. The epitaxial GaN stack includes stress compensation layers, a buried highly doped $n^+$ layer connected to the drain terminal to collect the current in a semi-vertical configuration, a relatively thick and lightly n-doped drift layer, a p-doped layer with different Mg chemical concentrations, and a top $n^+$ layer. The devices tested in this work vary in cleaning process, $SiO_2$ thickness, and magnesium chemical concentration.

## 3. Results and discussion

Fig. 2 shows $I_DV_{GS}$ comparison in semi-log scale for 2 devices with different cleaning processes: clean 1 and clean 2. The devices processed with clean 2 result in a higher $I_D$ in the transfer characteristic. In Figure 2 (b) the measured ON-resistance is compared for these devices, with different cleaning. In this work, the ON-resistance value is obtained by linear interpolation of the linear part of the $I_DV_{DS}$ curve performed at $V_{GS} = V_{TH} + 4$ V. The cleaning process has a significant influence on surface scattering and Coulomb scattering impacting the mobility and thus, the $R_{ON}$. As shown in Fig. 2 (b) the devices processed with clean 2 have a lower $R_{ON}$. The devices with thicker $SiO_2$ (Fig. 3 (a)) show higher threshold voltage but also higher ON-resistance. The devices fabricated with lower magnesium chemical concentration show lower threshold voltage and lower ON-resistance (Fig. 3 (b)). Considering the above-mentioned trends of the main device parameters, the best device performance is

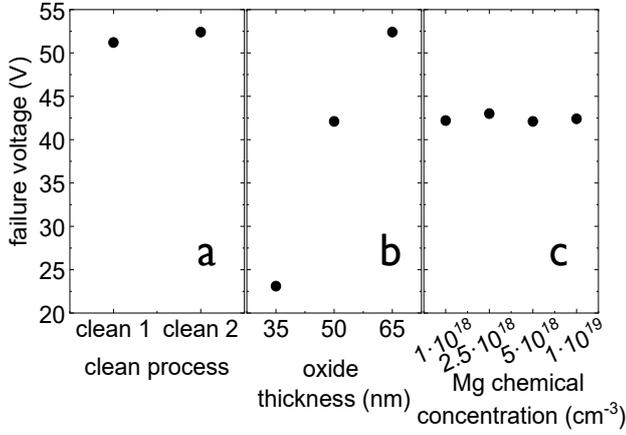

Fig. 5 Plot of the maximum failure voltage obtained for different cleaning processes (a), oxide thickness (b) and Mg chemical concentration (c).

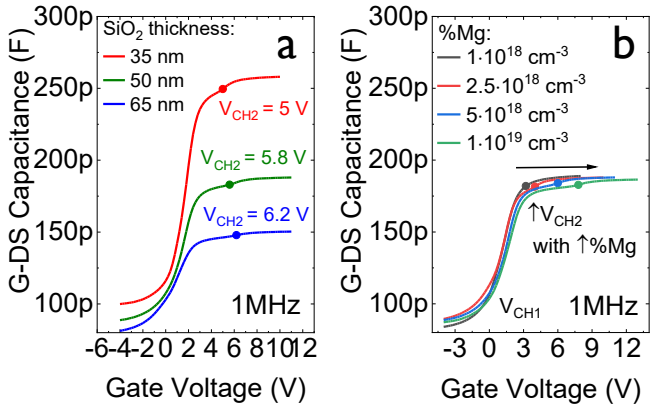

Fig. 6 (a) Comparison of the gate-drain capacitance curves for different $SiO_2$ thickness. (b) Comparison the gate-drain capacitance curves for different Mg chemical concentration of the p-GaN layer.

obtained for a gate dielectric consisting of 50 nm $SiO_2$ and for a p-body with a $2.5 \cdot 10^{18}$ cm$^{-3}$ chemical Mg concentration. In order to study the robustness of the gate stack, the gate-drain current has been investigated. We have already observed that in forward gate the gate-source and the gate-drain leakage curves are the same, due to the electron channel which surrounds the whole trench. In Fig.4 the trends of the $I_GV_{GS}$ (gate leakage) for devices with different cleaning process (Fig.4 (a)), oxide thickness (Fig.4 (b)), and chemical Mg concentration (Fig.4 (c)) are reported. The trend of the gate leakage shows that the failure occurs due to the breakdown of the gate dielectric. The failures (reported in Fig. 5) occur due to the breakdown of the gate dielectric. For all process splits, failure voltage is far beyond the maximum operating voltage (which is 12 V). The failure voltage increases by increasing the thickness of the $SiO_2$ (Fig. 4 and 5 (b)), and is independent from the gate trench cleaning, as well as the Mg chemical concentration of the p-GaN layer (Fig. 5 (a),(c)). The average critical electric field obtained with 50 nm and 65 nm $SiO_2$ is around 8 MV/cm, while it drops to 6 MV/cm for the 35 nm thick $SiO_2$ layer; this lower critical field is expected to originate from local roughness in the GaN sidewall on the gate trench, which will affect the non-uniformity and roughness of the deposited dielectric more, when a thinner layer is applied.

In Figure 6 the gate-drain-source (G-DS) capacitances (measured with the gate at one potential and with drain and source connected together to the other potential) of the devices under test are reported. The gate-drain-source capacitance shows two regions depending on the applied gate voltage. At low gate voltage ($V_{CH1}$) a rise in the capacitance is observed when the accumulation channel is formed at the interface between $Al_2O_3$ and n-GaN drift layer. At high gate voltage ($V_{CH2}$) a further bump in the capacitance is observed at the inversion channel formation. The maximum value of the capacitance is the sum of the gate oxide capacitance for the sidewalls and the bottom of the trench. In Figure 6 the comparison among gate-drain capacitance of devices with different $SiO_2$-thickness (a) and Mg chemical concentration (b) is reported. By increasing the $SiO_2$ thickness, the total capacitance (e.g. the oxide capacitance) decreases and the threshold voltage at which the formation of the inversion channel occurs ($V_{CH2}$) slightly increases: the same increase was observed in the device threshold voltage (see Figure 3 (a)). By increasing the Mg chemical concentration, the value of the total capacitance remains the same, however the threshold voltage at which the formation of the inversion channel occurs increases, in good agreement with the results reported from DC measurements (Figure 3 (b)) and with the model proposed in [29].

All the devices under test show a positive hysteresis in the transfer characteristic suggesting a trapping mechanism due to electron trapping.

$I_D$-$V_{GS}$ measurements are carried out between 0 V and $V_{TH}$ + 5V overdrive in forward (FWD) and backward (BWD) sweeping directions to further study the gate stability. The analysis was performed at 25 °C and at 150 °C for devices with a different cleaning process, $SiO_2$ thickness and Mg chemical concentration. The results of the analysis are reported in Fig. 7 (a,b,c). All devices show a positive hysteresis, which increases with temperature. However, by decreasing the Mg chemical concentration a decrease of the hysteresis is observed, especially at 150 °C. An estimate of the amount of trapped charge in the oxide is reported in Figure 10 (a).

The variation of the threshold voltage due to the trapped electrons in the oxide is given by [30]:

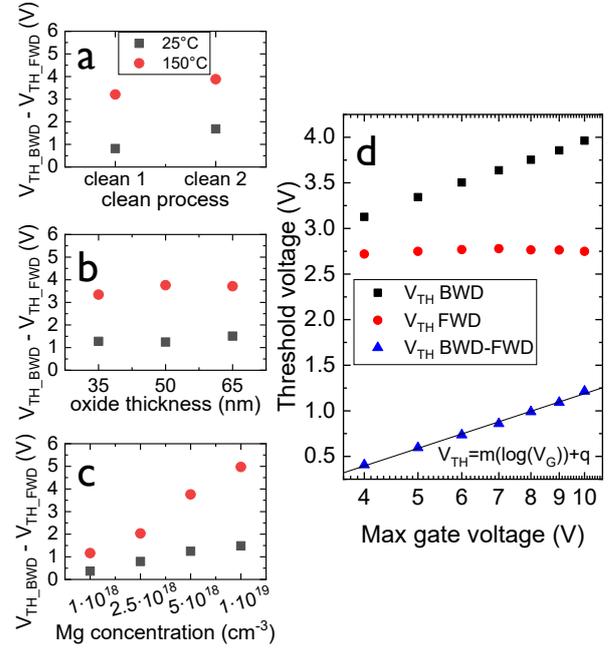

Fig. 7 Evaluation of the hysteresis ($V_{TH}$ BWD- $V_{TH}$ FWD) of the $I_DV_{GS}$ performed from 0 V to $V_{TH}$ + 5 V overdrive forward and backward performed at 25°C and at 150°C for devices with different cleaning processes (a), $SiO_2$ thicknesses (b) and Mg chemical concentrations (c). (d) Plot of the threshold voltage variation of the backward curve (black), of the forward curve (red) and of the hysteresis ($V_{TH}$ BWD-$V_{TH}$ FWD) of the $I_DV_{GS}$ curves performed forward and backward from 0 V to a maximum gate voltage increased from 4 V to 10 V with 1 V/step for the reference device.

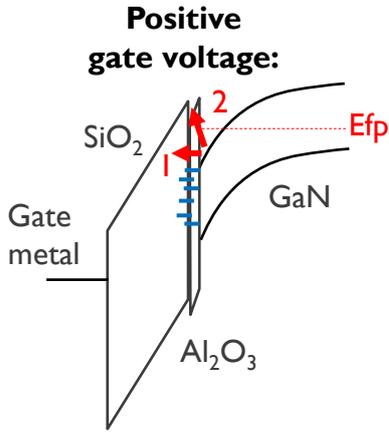

Fig. 8 Pictorial view of the two possible mechanisms responsible of the positive hysteresis: (1) trapping of electrons from the inversion channel to the Al$_2$O$_3$ and/or SiO$_2$ (2) electrons accumulate in the potential barrier at the Al$_2$O$_3$/SiO$_2$ interface.

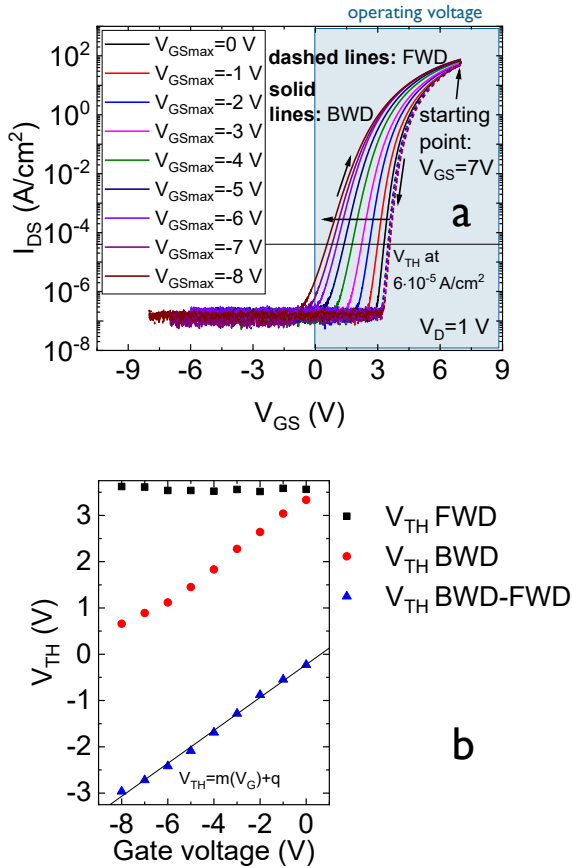

Fig. 9 (a) I$_D$V$_{GS}$ curves performed from 7 V to different negative voltages (0V, -1 V, -2 V, -3 V, -4 V, -5 V, -6 V, -7 V, -8V) (dashed lines) and backward (from negative voltages to 7 V (solid lines)). (b) Plot of the threshold voltage obtained from the I$_D$V$_{GS}$ curves performed in forward and in backward sweeping directions. A linear increase with the maximum applied negative voltage is observed.

$$\Delta V_{TH} = (V_{TH\ BWD} - V_{TH\ FWD}) = -Q_{ox} * C_{ox}$$

Where $Q_{ox}$ is the amount of net charge in the oxide (interface charge and oxide trapped charge), and $C_{ox}$ is the oxide capacitance,

$$C_{ox} = \frac{\varepsilon_0}{\left(\frac{T_{Al2O3}}{\varepsilon_{Al2O3}}\right) + \left(\frac{T_{SiO2}}{\varepsilon_{SiO2}}\right)}$$

In order to further analyze the trapping mechanism responsible for the positive hysteresis, several I$_D$V$_{GS}$ measurements (forward and backward) have been performed on the reference device (with 50 nm SiO$_2$ and 2.5·10$^{18}$ Mg chemical concentration) by increasing the maximum applied gate voltage. The plot of the variation of the threshold voltage is reported in Fig. 7 (d): the positive variation of the threshold voltage increases linearly with the logarithm of the maximum gate voltage. This result is consistent with those from metal-oxide-semiconductor field-effect transistors with high k dielectrics where $\Delta V_{TH}$ is observed to increase with a power law dependence on stress voltage [31][32]. The trapping mechanism is ascribed to injection of electrons from the channel toward the gate dielectric, the electrons can be trapped inside the Al$_2$O$_3$ and/or at the Al$_2$O$_3$/SiO$_2$ interface [33]. Moreover, a potential barrier of 1.47 eV [20] is created from the band alignment between SiO$_2$ and Al$_2$O$_3$, and electrons can accumulate at the interface between oxide layers. The semi-logarithmic trend of the threshold voltage with the maximum applied gate voltage can be explained by the fact that by increasing the applied gate voltage, the trapped and/or the accumulated electrons have a repulsive action towards other electrons and prevent them to be trapped/accumulated [34][35]. This trapping mechanism is the same in all the tested devices, independent from cleaning, SiO$_2$-thickness and Mg chemical concentration. By increasing the Mg concentration, the scattering effect at dielectric/GaN interface increases, possibly reducing mobility and increasing the trapping effect [26]. The observed trapping effect stays the same (Figure 4 (b)) by increasing the SiO$_2$-thickness suggesting that only the Al$_2$O$_3$ and/or the Al$_2$O$_3$-SiO$_2$ interface is affected by the electron trapping/accumulation. A pictorial representation of the trapping mechanisms responsible for the positive hysteresis is depicted in Figure 8.

The DUTs have also been investigated under negative gate bias by performing I$_D$V$_{GS}$ measurements from 7 V to a maximum gate negative voltage (upward) and from the same negative gate voltage to 7 V (backward). Although the device is biased well below the nominal gate operational voltage, this test is useful to further investigate the trapping phenomena occurring in the dielectric stack. In Figure 9 (a) the I$_D$V$_{GS}$ backward curves (negative to positive V$_{GS}$) are reported, the upward curves are overlapped. A negative shift of the threshold voltage is obtained as reported in Figure 9 (b) and it increases linearly with the maximum negative gate voltage applied.

The mechanism is possibly ascribed to the holes which are accumulating at the interface under negative bias and can be injected in the gate dielectric. The negative shift of the threshold voltage under negative bias is much larger than the positive shift observed under positive gate bias. This can be due to the fact that there is a lower potential barrier with the SiO$_2$. Long et al. [20] report that from the alignment of SiO$_2$ and GaN there are 2 eV of gap with the valence band and 3.6 eV with the conduction band) and more holes are injected compared to electrons in the FWD mode. In Figure 10 (b), the amount of the net charge in the oxide is calculated for each shift (V$_{TH}$ BWD-V$_{TH}$ FWD) of the threshold voltage when the device was submitted to negative gate bias.

**Conclusions**

In conclusion this work presents an exhaustive analysis of the optimization of the gate module in terms of cleaning process, gate dielectric thickness and body layer doping level. The optimum device is selected with 50 nm of $SiO_2$ as gate dielectric and $2.5 \cdot 10^{18}$ of Mg chemical concentration in order to have better gate yield, higher gate robustness and reduced $R_{ON}$.

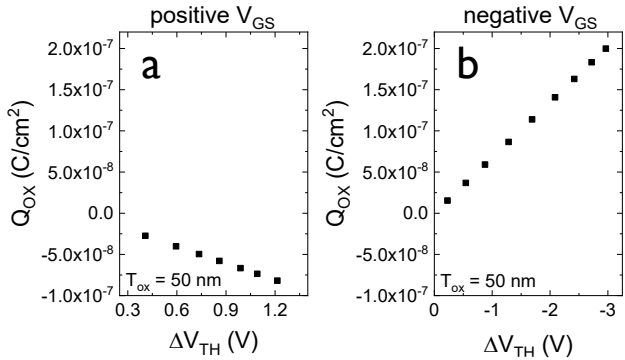

Fig. 10 Net charge in the oxide ($SiO_2$+$Al_2O_3$) calculated for each variation of the threshold voltage under positive gate bias (a), and negative gate bias (b).

The gate stability under positive and negative gate bias is analyzed: presumably, electrons injected from the inversion layer are trapped into the dielectric and/or at the $Al_2O_3$/$SiO_2$ interface inducing a positive shift of the threshold voltage not depending on the thickness of the $SiO_2$. The trapping effect is reduced for lower Mg concentrations, probably due to reduced scattering. By applying a negative gate voltage, a negative shift of the threshold voltage is induced. This is ascribed to the trapping of holes.

**Acknowledgements**

This paper has received funding from the ECSEL Joint Undertaking (JU) under grant agreement No 826392. The JU receives support from the European Union's Horizon 2020 research and innovation programme and Austria, Belgium, Germany, Italy, Norway, Slovakia, Spain, Sweden, Switzerland.